\documentclass[journal=jacsat,manuscript=article,layout=twocolumn]{achemso}

\usepackage[version=3]{mhchem} 
\usepackage{textcomp}
\usepackage{natbib}
\usepackage{color}

\newcommand*{\We}{\mathrm We}

\def\be{\begin{equation}}
\def\ee{\end{equation}}

\author{Hrudya Nair}
\affiliation{Physics of Fluids, University of Twente, P. O. Box 217, 7500 AE Enschede, The Netherlands}
\alsoaffiliation{Catalytic Processes and Materials, University of Twente, P. O. Box 217, 7500 AE Enschede, The Netherlands}
\alsoaffiliation{MESA+ Institute for Nanotechnology, University of Twente, P. O. Box 217, 7500 AE Enschede, The Netherlands}
\altaffiliation{These authors contributed equally to this work.}
\author{Hendrik J. J. Staat}
\affiliation{Physics of Fluids, University of Twente, P. O. Box 217, 7500 AE Enschede, The Netherlands}
\altaffiliation{These authors contributed equally to this work.}
\author{Tuan Tran}
\affiliation{Physics of Fluids, University of Twente, P. O. Box 217, 7500 AE Enschede, The Netherlands}
\altaffiliation{Present address: School of Mechanical and Aerospace Engineering, Nanyang Technological
University, 50 Nanyang Avenue, Singapore 639798}
\email{ttran@ntu.edu.sg}
\author{Arie van Houselt}
\affiliation{Catalytic Processes and Materials, University of Twente, P. O. Box 217, 7500 AE Enschede, The Netherlands}
\alsoaffiliation{MESA+ Institute for Nanotechnology, University of Twente, P. O. Box 217, 7500 AE Enschede, The Netherlands}
\author{Andrea Prosperetti}
\affiliation{Physics of Fluids, University of Twente, P. O. Box 217, 7500 AE Enschede, The Netherlands}
\alsoaffiliation{Department of Mechanical Engineering, Johns Hopkins University, Baltimore, MD 21218, USA }
\author{Detlef Lohse}
\affiliation{Physics of Fluids, University of Twente, P. O. Box 217, 7500 AE Enschede, The Netherlands}
\alsoaffiliation{MESA+ Institute for Nanotechnology, University of Twente, P. O. Box 217, 7500 AE Enschede, The Netherlands}
\email{d.lohse@utwente.nl}
\author{Chao Sun}
\affiliation{Physics of Fluids, University of Twente, P. O. Box 217, 7500 AE Enschede, The Netherlands}
\alsoaffiliation{MESA+ Institute for Nanotechnology, University of Twente, P. O. Box 217, 7500 AE Enschede, The Netherlands}
\email{c.sun@utwente.nl}
\title{Leidenfrost temperature increase for impacting droplets on carbon-nanofiber surfaces}

\begin{document}

\begin{abstract}
\small{
Droplets impacting on a superheated surface can either exhibit a 
contact boiling regime, in which they make direct contact with the surface 
and boil violently, or a film boiling regime, in which they remain separated 
from the surface by their own vapor. The transition from the contact to the 
film boiling regime depends not only on the temperature of the surface and 
kinetic energy of the droplet, but also on the size of the structures 
fabricated on the surface. Here we experimentally show that 
surfaces covered with carbon-nanofibers delay the transition to film boiling 
to much higher temperature compared to smooth surfaces. We present 
physical arguments showing that, because of the small scale  
of the carbon fibers, they are cooled by the vapor flow just before the 
liquid impact, thus permitting contact boiling up to much higher temperatures 
than on smooth surfaces. We also show that, as long as the 
impact is in the film boiling regime, the spreading factor of impacting 
droplets follows the same $\We^{3/10}$ scaling (with $\We$ the Weber number) 
found for smooth surfaces, which is caused by the vapor flow underneath the 
droplet. 
}
\end{abstract}


\section{Introduction}


Spray cooling is an effective heat transfer mechanism as it is capable of 
delivering spatially uniform and high heat transfer 
rates (see e.g.\cite{kim07,agostini07,EbadianLin11}). An important new 
application of this technology is in electronic cooling, where the growing 
power consumption and decreasing sizes pose increasingly challenging heat 
dissipation demands (see e.g. \cite{pautsch05,VisariaMudawar09}). Other 
common situations in which cold drops impact hot surfaces are found in 
internal combustion engines (see e.g. \cite{Arcoumanisetal98,PanaoMoreira09}),
quenching of aluminum and steel (see e.g. \cite{MascarenhasMudawar12}), 
fire suppression (see e.g. \cite{Yoonetal10,ChenWang11}) and others. 

In all these applications a stream of fine droplets dispensed, e.g., from a 
nozzle impinges on a solid surface and cools it by a combination of 
sensible heat absorption and latent heat of vaporization. 
Due to the inherent complexity of the phenomenon and the large number of 
parameters involved, such as droplet size, velocity distribution, droplet 
number density and material properties, many aspects of the 
physical mechanisms involved still remain incompletely 
understood\cite{kim07,moreira10,Berberovicetal11}. 

A fundamental understanding of the impact of an individual droplet on 
superheated surfaces is the first step toward a better understanding and 
eventual optimization of the process. Various aspects of this particular 
problem have been investigated, such as the effect of droplet size, 
velocity, physical properties 
(see e.g. \cite{Yarin06,Herbertetal13}), and surface roughness 
(see e.g. \cite{bernardin97,tran13}),
the transition between different boiling regimes 
(see e.g. \cite{bernardin99,wang00,bernardin04,tran12,tran13}),
the surface temperature change and heat transfer during 
impact (see e.g. \cite{bernardin97,Leeetal01,weickgenannt11,weickgenannt11b}),
the residence time of the impacting droplet (see e.g. \cite{chen07,tran13}), 
the spreading factor (see e.g. \cite{chen07b,tran12,tran13}) and others. 

An important quantitative feature of the phenomenon is the 
transition temperature $T_L$ between the contact boiling regime, where
the liquid makes direct contact with the heated surface, and 
the film boiling regime, where a stable vapor layer between the 
liquid and the surface is formed during impact.
As the rate of heat transfer in the film boiling regime 
is significantly reduced  
due to the poor thermal conductivity of the vapor layer, 
this regime should be avoided for applications that require
high heat transfer rates. Methods to increase $T_L$, or delay the onset of 
the film boiling regime, are therefore of great interest for such 
applications. 

Recently, surfaces covered with nanofibers 
were shown to effectively enhance the heat transfer from 
the surface to a liquid in contact with it\cite{weickgenannt11b,jun2013pool}.
In particular, it was reported that for impacting 
ethanol droplets on surfaces covered with {\it nano}fiber mats, 
the film boiling behavior 
was not observed even when the surface temperature was as high 
as 300$\,^\circ$C\cite{weickgenannt11}, which implies that
the transition temperature to film boiling is {\it increased}
compared to that observed on smooth surfaces. 
This is in marked contrast with the impact on surfaces covered with 
{\it micro}structures, for which the transition temperature is considerably {\it decreased} compared to a smooth surface \cite{tran13}. 
Indeed, numerous questions regarding 
the effects of nanostructures on the 
transition temperature are still open. 
First of all, why do nanofibers cause a higher $T_L$ compared to that on 
smooth surfaces? And, further, what is the transition 
temperature $T_L$ on this type of surfaces? how does it change with the 
size of the nanostructures on the surfaces?

To answer these questions, in this paper we study the impact of droplets 
on heated surfaces covered with carbon nanofibers (CNF), 
which are carbonaceous structures 
grown by catalytic vapor deposition of hydrocarbons. 
This type of nanostructures is well-known for their unique 
physical and chemical properties with a tunable morphology 
(the diameter can be varied from a few to hundreds of nanometers, 
the height can be controlled from a few micrometers to millimeters),
which in turn can be exploited for tuning 
the roughness, porosity, and surface area\cite{bitter10}. 

We use two types of 
CNF surfaces corresponding to two different typical fiber lengths and a 
smooth silicon surface. 
For each type of surface, 
we determine the transition temperature and its dependence 
on the impact velocity. 
We propose a quantitative explanation of the effect of nanofibers 
on the transition temperature $T_L$. Furthermore, for impact of droplets 
in the film boiling regime, we measure the spreading factor and compare 
our data with existing models.

%
%
%
%

\section{Experimental details}

\subsection{Synthesis of carbon nanofiber layers}

\begin{figure}
\begin{center}
\includegraphics[width=8.5cm]{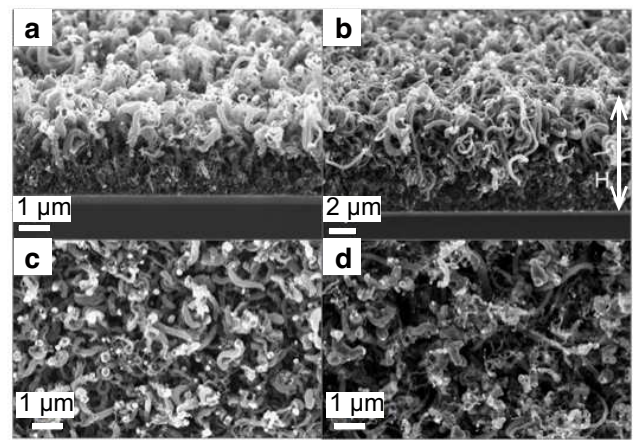}
\caption{\small{Scanning electron microscope (SEM) images 
showing  side views of the carbon nanofiber (CNF) layers with a 
synthesis time of (a) 11$\,$min and (b) 14$\,$min. 
The arrow in (b) indicates the height H of the CNF layer. 
The corresponding top-view SEM images are shown in 
(c) for a synthesis time of 11$\,$min and 
in (d) for a synthesis time of 14 min.  
The bar represents 1$\,$\textmu m in (a), (c) and (d), 
and 2$\,$\textmu m in (b). 
}}
\label{sem}
\end{center}
\end{figure}

\begin{figure} 
\begin{center} 
\includegraphics[width=8.5cm]{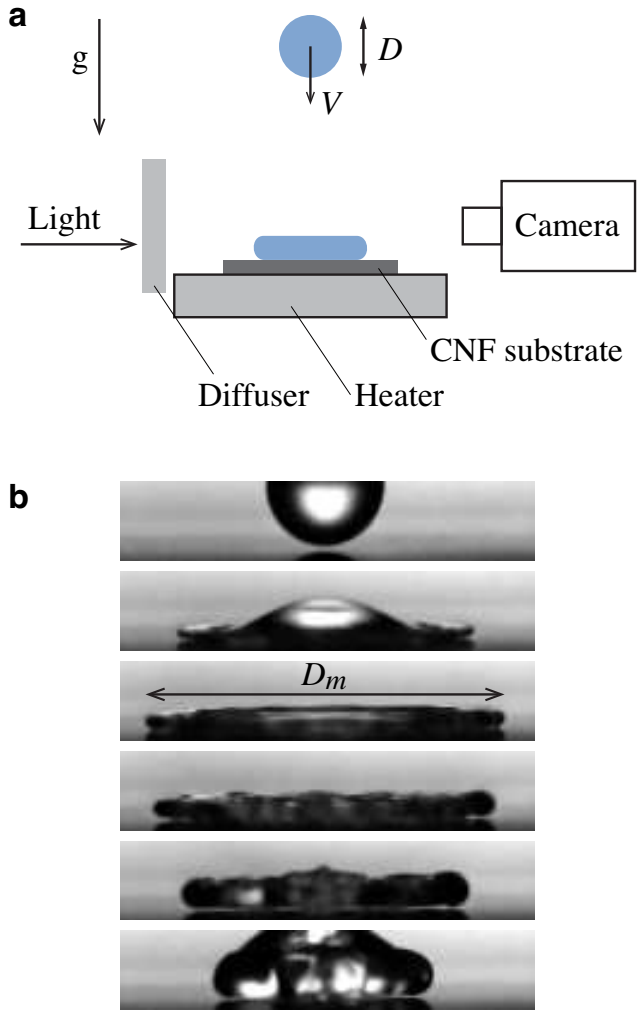} 
\caption{\small{
(a) Schematic (not to scale) of the experimental setup
used to observe the characteristic behaviors of impacting 
droplets on heated surfaces. The surface of interest is placed on a 
heater, which can be heated up to 500$\,^\circ$C. 
FC-72 droplets of diameter $D$ impact the heated sample
with impact velocity $V$. The behavior of the impacting droplets 
is recorded from the side by a high-speed camera (Photron SA1.1). 
From the recordings, $D$, $V$, and the maximum spreading $D_m$ of the droplet 
can be measured.  
(b) Series of snapshots of an impacting droplet in the film boiling regime
showing how $D_m$ is measured as the maximum 
horizontal extension of the droplet.
}} 
\label{setup} 
\end{center} 
\end{figure} 

Carbon nanofibers (CNFs) were synthesized on oxidized silicon wafers 
(p-type, $5-10\,$Ohm-cm resistivity, 
$100\,$mm diameter, $525 \pm 25\, $\textmu m 
thickness, $\{100\}$ crystal orientation; Okmetic Finland)
 using nickel (Ni) thin film as catalyst. 
 First, a \ce{SiO2} layer of 220$\,$nm thickness was grown via wet oxidation 
 (45$\,$min, $1000\,^\circ$C) on these silicon substrates. 
 Second, a  pattern was defined in spin-coated photoresist 
 (Olin, 906-12), resulting in unmasked squares of 8$\,$mm$\times$8$\,$mm, 
 by means of standard UV lithography (EVG 620). 
 Further, a 10$\,$nm tantalum layer followed by a 25$\,$nm nickel layer 
 was deposited via electron-beam evaporation. 
 Finally the samples were subjected to an ultrasonic lift-off step in acetone 
 ($>20\,$min; VLSI 100038, BASF), followed by rinsing in water and spin drying. 
 These nickel-coated substrates were diced into 1$\,$cm$\times$1$\,$cm samples 
 (Disco DAD-321 dicing machine). 
 To remove  organic contaminants, 
 these samples were ultrasonically cleaned in acetone 
 (10$\,$min, Branson 200 ultrasonic cleaner) 
 and de-ionized water (2$\,$min, 25$^\circ$C). \cite{nair12}

 After drying with synthetic air, 
 the samples were placed centrally on a 
 flat quartz boat inside a quartz reactor and were loaded into a horizontal 
 oven equipped with three temperature controllers along it. 
 Nitrogen (\ce{N2}; 99.999$\,$\%, INDUGAS NV.) 
was used as carrier gas during heating, 
pretreatment, CNF synthesis and cooling. 
First, the temperature was increased ($5\,$K$\,$min$^{-1}$) to 500$\,^\circ$C. 
Second, the samples were pretreated with 20 vol.$\,$\% 
of hydrogen (\ce{H2}; 99.999$\,$\%, INDUGAS NV.) for 2$\,$hours at a total flow 
rate of 50$\,$ml$\,$min$^{-1}$ in order to reduce the passivated \ce{Ni} thin film. 
Subsequently the temperature was increased ($5\,$K$\,$min$^{-1}$) to 635$\,^\circ$C, 
at which temperature the CNF synthesis was performed via 
catalytic vapor decomposition using  
25 vol.$\,$\% ethylene (\ce{C2H4}; 99.95$\,$\% Praxair Inc.) 
and 6.25 vol.$\,$\% \ce{H2} in a total flow rate 100$\,$ml$\,$min$^{-1}$. 
Finally the samples were cooled down to room temperature ($10\,$K$\,$min$^{-1}$). 

Two sets of samples were used for the droplet impact studies. 
One set was obtained after a CNF synthesis time of 11$\,$min, 
resulting in a CNF layer thickness 3.4$\pm$0.3$\,$\textmu m. 
The other set was obtained after a CNF synthesis time of 14$\,$min, 
resulting in a CNF layer thickness of 7.5$\pm$0.7$\,$\textmu m. 
More details of the influence of synthesis time on CNF layer thickness have been reported previously 
\cite{nair12}. 
These samples will be termed as CNF(3.5) and CNF(7.5), respectively.
\ref{sem} shows representative scanning electron microscope (SEM) images 
with the side views (\ref{sem}(a) and (b)) 
and top views (\ref{sem}(c) and (d)) 
of the surfaces CNF(3.5) and CNF(7.5).
  
The thickness of the CNF layers was determined using 5 
representative cross-sectional SEM images taken at various positions 
on the sample (10 height measurements were averaged per SEM image). 
The width of the nanofibers ranges from 32$\,$nm to 220$\,$nm
with average value of 127$\,$nm.

\subsection{Experimental method} 
\begin{figure*} 
\begin{center} 
\includegraphics[width=14cm]{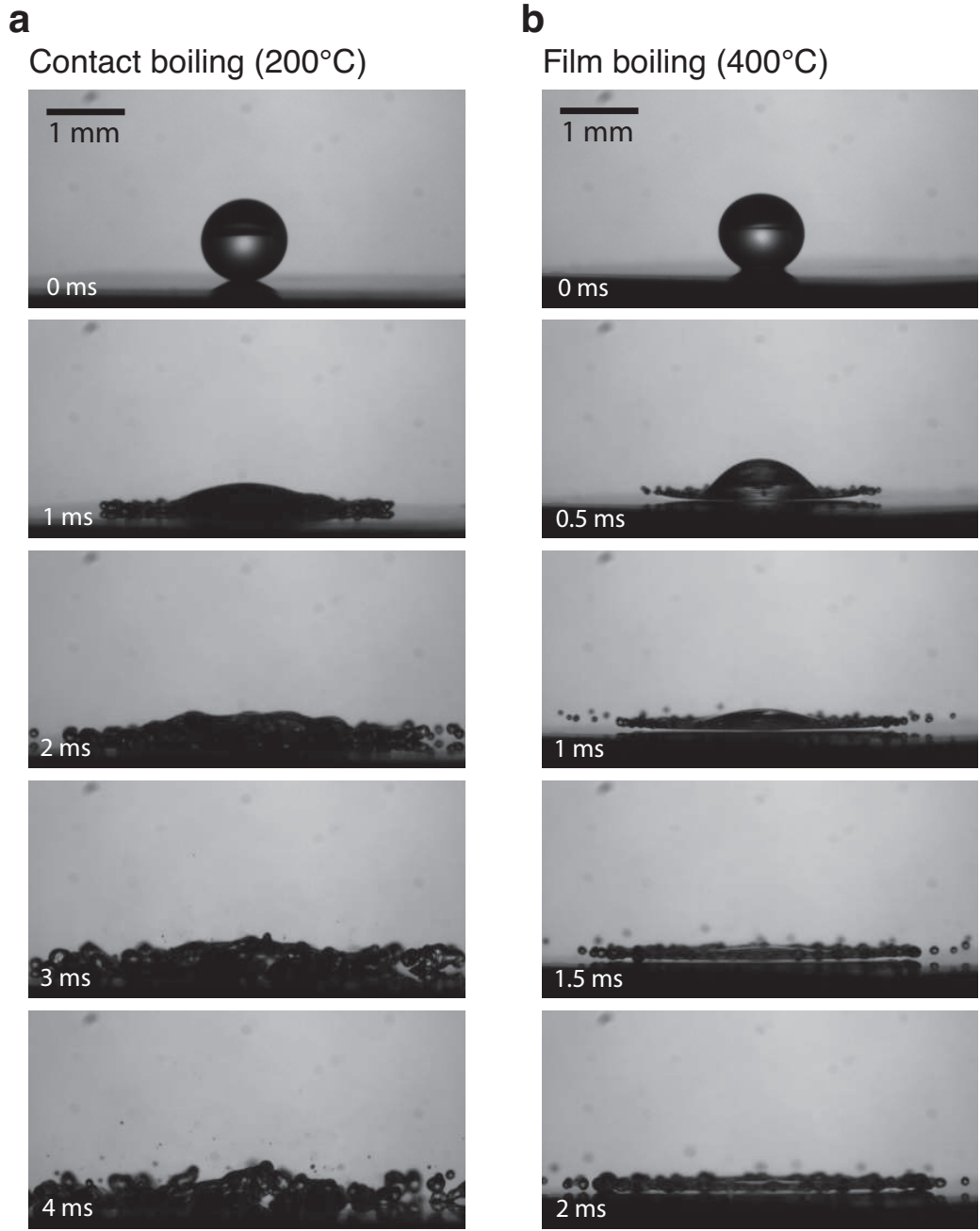} 
\caption{\small{(a) 
Representative images showing the characteristic boiling behavior 
of an impacting FC-72 droplet on a 7.5 {\textmu}m-thick CNF  surface
 in the contact boiling regime, $T=200\,^\circ$C. 
The diameter of the impacting droplet is $D=1.1\,$mm, the impact velocity 
$V=1.0\,$m/s and the Weber number ${\rm We} = 154$.
(b) Representative images of a FC-72 droplet with the same diameter 
and velocity impacting on the same surface as in (a), but at the higher 
surface temperature, $T=400\,^\circ$C. In this case, the impact is in the 
film boiling regime.   
}} 
\label{series} 
\end{center} 
\end{figure*}

A schematic diagram of the experimental setup is shown 
in  \ref{setup}. 
All droplet impact experiments were performed with 
FC-72 (3M Fluorinert Electronic Liquid), 
a dielectric fluid commonly used in 
electronics cooling applications.
The liquid has boiling point $T_b = 56\,^\circ$C, 
density $\rho_l= 1680\,$kg$\,$m$^{-3}$, and
surface tension $\sigma =0.01\,$N$\,$m$^{-1}$.
We generate droplets by
using a syringe pump (PHD 2000 Infusion, Harvard Apparatus)
to inject liquid into a small fused silica needle 
where the droplets are formed at the tip.
The flow rate is kept at a small value ($\approx 0.1\,$mL$\,$min$^{-1}$) 
so that droplet detachment from the needle is due only to 
gravitational force, hence keeping the droplet size uniform.
After detaching from the needle, a droplet falls on the target surface
placed on a brass plate with a cartridge heater 
and a thermocouple (Omega Inc.) embedded inside.
The surface temperature $T$ was set by a controller
and was varied between 
$60\,^\circ {\rm C}$ and $450 \,^\circ {\rm C}$. 
This temperature was also 
measured independently by a surface temperature probe 
(Tempcontrol B.V.). 
The difference between the controller's set point and the surface
probe measurement was less than $3 \,{\rm K}$.
Thus we take the controller's set point as 
the surface temperature $T$ of the surface.

Recordings of the impact events 
were made with a high-speed camera (Photron SA1.1) 
(see \ref{setup}).
From these high-speed recordings, 
the boiling behaviors were analyzed, 
and the droplet diameter $D$, the impact velocity $V$ and 
the maximum spreading diameter $D_m$ (see \ref{setup}) were measured. 
From the measured diameter and velocity, we calculated 
the Weber number ${\rm We} = \rho_l D V^2/\sigma$, which 
is a dimensionless number that characterizes the 
droplet's kinetic energy compared to its surface energy.
The impact velocity $V$ was varied by changing the needle's height.
Impact events were repeated at least three times for every combination 
of $V$ and $T$ to test reproducibility of the experiment.

\subsection{Characterization of boiling behavior} 
By varying the surface temperature between 
$60\,^\circ{\rm C}$ and 450$\,^\circ$C 
and the Weber number between 10 and 1000, we observed 
two characteristic boiling behaviors: contact boiling and film boiling.
In \ref{series} we show two series of images 
to illustrate the difference between these two regimes. 
The essential difference between the two is 
whether or not the liquid makes direct contact with the heated surface
during impact \cite{fujimoto10,tran12,quere13}. 
In the contact boiling regime (\ref{series}(a)), 
as the pressure of the vapor generated 
underneath the droplet is 
not sufficient to support the droplet's dynamic pressure, 
the liquid touches the heated surface and quickly boils due to 
the high heat flux through the contact area. The recorded snapshots show
the small droplets ejected as a result of the boiling process. 
In contrast, an impacting droplet in the film boiling 
regime is separated from the heated surface by a developing vapor layer (see 
\ref{series}(b)).
This vapor layer insulates the droplet during the impact time,
hence prevents the liquid from boiling violently. 

By carefully analyzing the recorded movies of impacting droplets, 
we categorized the impact as being in the film boiling 
regime when droplet ejection or vapor bubble generation were not observed.

%
%
%
 
\section{Results and discussions}
 
 \subsection{Dynamic Leidenfrost temperature}
 \begin{figure} 
\begin{center} 
\includegraphics[width=8.6cm]{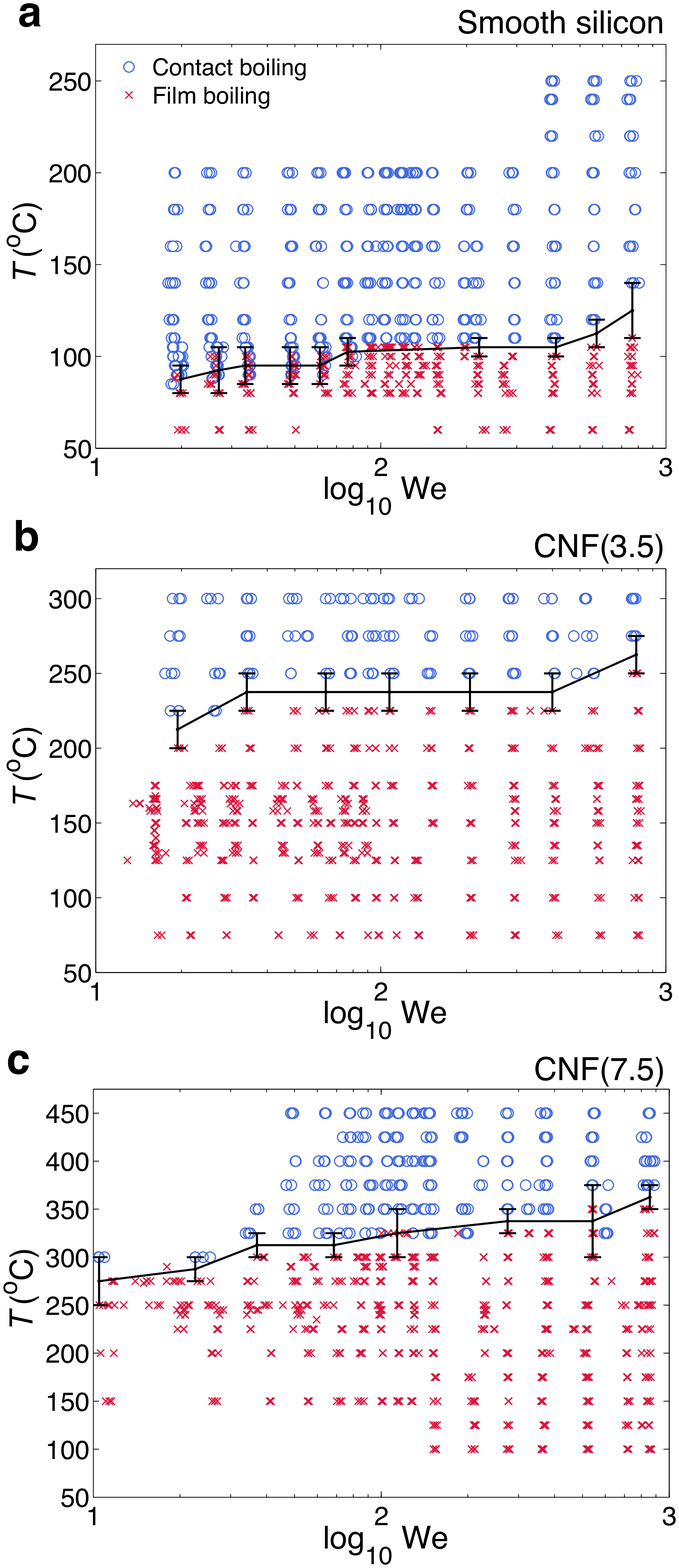} 
\caption{\small{(a) Phase diagram 
showing the characteristic boiling behaviors of impacting FC-72 droplets on 
smooth silicon surfaces. The contact boiling regime (red crosses) and film 
boiling regime (open blue circles) are separated by a transition band, 
indicated by the vertical bars, where both characteristic behaviors were 
observed. (b) Phase diagram for surfaces covered by a 3.5 {\textmu}m-thick 
CNF layer.  (c) Phase diagram for surfaces covered by a 7.5 {\textmu}m-thick 
CNF layer. Note the much larger temperature ranges in (b) and (c).
}} 
\label{pd} 
\end{center} 
\end{figure} 

 In \ref{pd} we show phase diagrams of the  characteristic boiling behavior 
of impacting droplets on  smooth silicon surfaces, CNF(3.5), and CNF(7.5), 
respectively. 
The temperature 
ranges were  $60\,^\circ{\rm C}$ to $250\,^\circ{\rm C}$ for the smooth 
silicon surfaces, $60\,^\circ{\rm C}$ to $300\,^\circ{\rm C}$ for the CNF(3.5) surfaces, and  $100\,^\circ$C to 450$\,^\circ$C for the CNF(7.5) 
surfaces. 
   In each phase diagram, there is a clear transition between the 
 contact and the film boiling regimes. This transition temperature is marked 
 by a solid line, with the vertical bars indicating the intermediate region 
where both boiling behaviors were observed. 
 The transition temperature, known as the dynamic Leidenfrost temperature 
$T_L$, increases with increasing kinetic energy of impacting droplets.
  This dependence of $T_L$ on ${\rm We}$ is qualitatively similar to that
  found previously for droplets impacting on smooth and 
micro-structured surfaces \cite{bernardin97,tran12,tran13},
and is expected: the increasing momentum of the impact forces the droplet 
into contact with the surface at larger and larger temperature. 

\begin{figure} 
\begin{center} 
\includegraphics[width=8.4cm]{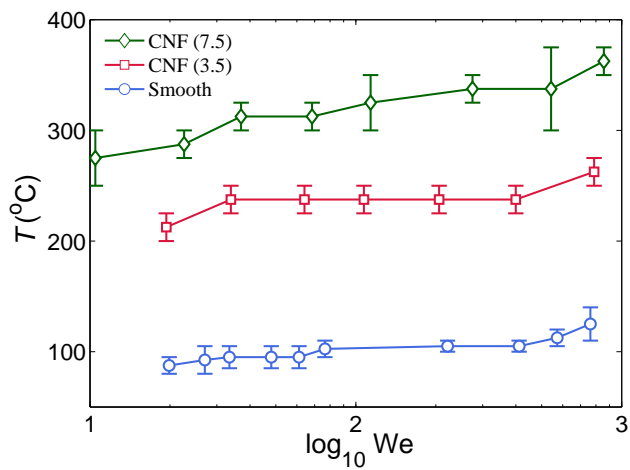} 
\caption{\small{Dynamic Leidenfrost temperature (transition from contact to 
film boiling) for smooth silicon, and surfaces covered by a 
3.5 {\textmu}m- and a 7.5 {\textmu}m-thick layer of carbon nanofibers.  
}} 
\label{T_L} 
\end{center} 
\end{figure}

These results, however, are in stark contrast with those found for smooth 
and microstructured surfaces in two respects, as can be seen from  
 \ref{T_L} in which $T_L$ for the smooth and CNF surfaces is compared.  
The first unexpected finding is that, while $T_L$ is lower for 
microstructured surfaces as compared with smooth ones \cite{tran13},
it is actually 
higher in the case of carbon nanofibers. For example, for ${\rm We}=100$, 
$T_L$ for the smooth surface is 110$\,^\circ$C, whereas for  CNF(3.5)  
and CNF(7.5) it increases to 250$\,^\circ$C and 350$\,^\circ$C, 
respectively.  Secondly, $T_L$ increases with nanofiber length, again in 
contrast with surfaces covered with micrometer-size pillars 
for which, for given shape and spacing, the microstructure height is 
inversely correlated with $T_L$ \cite{tran13}. The tentative explanation 
of that  latter finding offered in Ref. \cite{tran13} is that the surface of the 
impacting liquid tends to penetrate the space between the pillars. This 
causes the liquid surface area to increase, the more the 
higher the pillars. As a consequence, the vapor generation rate also increases 
and the film boiling regime sets in  at a lower temperature. 

As explanation of the opposite behavior found with carbon 
nanofibers we suggest  that they are efficiently cooled by the vapor flow before the 
drop touches the CNF surface. To support this conjecture in the following 
subsection we will estimate the time scale $\tau_c$ 
for the temperature of the nanofibers to cool, and compare it with the time scale $\tau_e$ the nanofiber is 
exposed to the vapor flow (which will be found to be somewhat larger), 
and also with the time scale $\tau_h$ for the heat flow 
inside the nanofiber (which will be  found to be much larger). 

\subsection{Estimate of the relevant time scales}
We start with the estimate for the time scale $\tau_c$ for the cooling of the nanofibers
by the ``vapor wind''.
Since the cross section of the nanofibers is of the order of 100 nm, the time scale  
$\tau_c$  can be estimated by assuming their temperature  to be 
uniform, which is the  so-called lumped capacitance approximation (see 
e.g.~\cite{Incroperaetal}). This time scale can then be estimated to be 
\begin{equation}
\tau_c\sim {w \rho_n C_n\over h}, 
\end{equation}
in which $w$ is the diameter of the nanofiber, 
$\rho_n$ and $C_n$ its density and specific heat, and $h$ the heat 
transfer coefficient. The latter can be expressed in terms of a Nusselt 
number, ${\rm Nu}= w h/K_v$, with $K_v$ the vapor thermal conductivity, 
so that 
\begin{equation}
 \tau_c\,=\, \frac{\rho_n C_n w^2}{K_v {\rm Nu}}
\,=\, \frac{\rho_n C_n}{\rho_v C_v} \frac{w^2}{\kappa_v {\rm Nu}}\, ,
\label{eqtauc}
\end{equation}
in which $\rho_v$, $C_v$, and $\kappa_v$ are the vapor density, specific heat, and 
thermal diffusivity, respectively. The (approximate) proportionality of $\tau_c$ to the 
square of the 
fiber size is a particularly noteworthy feature of this expression. 
In standard correlations (see e.g.~\cite{Incroperaetal}), Nu is given as 
a function of the fluid Prandtl and Reynolds numbers. No measured value 
for the former seems to be available for FC-72 vapor, but it is well known 
that the Prandtl number of gases is close to 1 and we can safely use this 
estimate here. Estimation of the Reynolds number requires a value for the 
viscosity of the vapor which, again, does not seem to have been measured. The 
order of magnitude of the viscosity of many gases and vapors is 10$^{-5}$ Pa s, 
and this is the value we will use. 
The density of FC-72 vapor at the boiling point 56 $^\circ$C is about 
11.5 kg/m$^3$. Taking $w \sim$ 100 nm and using these values we then find 
${\rm Re}\sim 0.1\,V_v$, with $V_v$ the vapor velocity in m/s. This 
quantity has been estimated in \cite{tran13} (see equation (13) of that paper) where it was found to be of the 
order of
\begin{equation}
 V_v \,\sim \,\left(\frac{\rho_l C_v \Delta T}{\rho_v L {\rm Pr}_v}\right)
^{1/2}V\,,
\label{evpvl}
\end{equation}
with $\rho_l$ the liquid density, $\Delta T$ the liquid-surface temperature 
difference, $L$ the latent heat and ${\rm Pr}_v$ the vapor Prandtl number. 
With $\rho_l$ = 1680 kg/m$^3$, $L$ = 88 kJ/kg, $C_v$ =  910 J/kg K and 
$\rho_v$ = 11.5 kg/m$^3$ (values at 56 $^\circ$C), $\Delta T \sim$ 100$\,$K, 
$V$ the impact velocity $\sim$ 1 m/s and again taking ${\rm Pr}_v\sim$ 1, we find 
$V_v\sim$ 12 m/s so that ${\rm Re} \sim 1.2$. The 
Churchill-Bernstein correlation ~\cite{Incroperaetal} then gives a Nusselt 
number of about 1. Use of equation (\ref{eqtauc}) requires values of $K_v$ or $\kappa_v$, 
neither of which seems to be available. For many gases and vapors $\kappa_v$ is of 
the order of 10$^{-5}$ m$^2$/s. With this estimate, taking $\rho_n \simeq$ 
2267 kg/m$^3$, $C_n \simeq$ 709 J/kg K and, again, $w \sim$ 100 nm, we find 
from eq.\ (\ref{eqtauc}) $\tau_c \simeq$ 150 ns. 

This time scale has to be compared with  the characteristic time $\tau_e$ during which the fiber 
is exposed to the cooler vapor until the liquid makes contact with it, which  
can be estimated as \begin{equation}
\tau_e \sim {H_v\over V}, \end{equation} where $H_v \sim D {\rm St}^{-2/3}$ 
is the characteristic thickness of the vapor layer at which the drop starts 
being deformed due to the increasing pressure on its underside 
\cite{mandre09}.  Here, as above,  $D$ is the droplet diameter, $V$ is the impact velocity, 
 and  ${\rm St} = \rho_l V D /\mu_v$ is the Stokes number, where 
 $\mu_v$ is the viscosity of vapor. Hence we obtain the time during which 
 the nanofibers are exposed to the cooler vapor flow 
$\tau_e \sim D {\rm St}^{-2/3} /V$. In the use of this estimate we again 
encounter the problem that $\mu_v$ is not available but, if we use the same
 estimate $\mu_v \sim 10^{-5}$ Pa$\,$s as before and take 
  $V\simeq$ 1 m/s, $D \simeq$ 1\,mm, 
 we find $\tau_e \simeq$ 330$\,$ns, which is seen to be long enough to cause 
a substantial cooling of the fibers.

Of course, as the fibers are cooled by the vapor, heat flows towards their tips
from the silicon substrate with a characteristic time 
\begin{equation}
\tau_h= {\ell^2\over \kappa_c},\end{equation} in which $\ell$ is the fiber length and $\kappa_c$ its 
thermal diffusivity of the carbon nanofibers. Since, in this experiment, the fibers had 
not been heat-treated, we can estimate their thermal conductivity on the basis 
of the results of Ref.~\cite{MayhewPrakash13} as $K_c$ = 4.6 W/m K and, 
therefore, $\kappa_c\sim 2.86\times 10^{-6}$ m$^2$/s. For the shorter fibers 
$\ell \simeq $ 3.4 {\textmu}m and, therefore, $\tau_h \sim$ 4 {\textmu}s
while, for the longer fibers,  $\ell \simeq $ 7.5 {\textmu}m and 
$\tau_h \sim$ 20  {\textmu}s. These times are much longer than both the 
cooling time and the exposure time to the vapor flow, which implies that 
the liquid encounters fibers at a much cooler temperature than the core silicon 
substrate. This circumstance would explain why the CNF surfaces 
require a higher temperature to achieve the film boiling regime compared to 
the smooth surfaces, and why the transition temperature increases with
the fiber length. 

The size of the cross section of the fibers in our experiment is close to 
the cross-over value at which cooling and exposure to the vapor flow have
comparable time scales. It follows that fibers or, more generally, 
microstructures with a larger cross section would be insensitive to the 
cooling effect. As a check of this expectation we can apply the same 
estimates to the case of the microstructured 
surfaces studied earlier \cite{tran13}. In that case the fluid was water for 
which, of course, all the required physical properties are well known. The 
microstructures had the form of silicon pillars with a square cross section 
of about 10$\times$10 {\textmu}m$^2$ and heights from 2 to 8 {\textmu}m. The 
vapor velocity estimated from eq.\ (\ref{evpvl}), again with $\Delta T \sim 100K 
$ and $V\sim$ 1 m/s, is found  to be $V_v\sim$ 12 m/s. The 
corresponding Reynolds 
number is Re$\sim$ 6 with a corresponding Nusselt number Nu$\sim$ 1.7. 
In this case $\rho_n$ = 2330 kg/m$^3$, $C_n$ = 705 J/kg K and eq.\ (\ref{eqtauc}) 
gives $\tau_c\sim$ 6.6 ms. The exposure time to the vapor is not very 
different from the previous estimate, and is therefore several orders of 
magnitude shorter. It is evident that, in this case, the vapor flow is 
just a small perturbation which does not have an appreciable effect on
the pillar temperature.  

 \subsection{Spreading factor}

\begin{figure} 
\begin{center} 
\includegraphics[width=8.5cm]{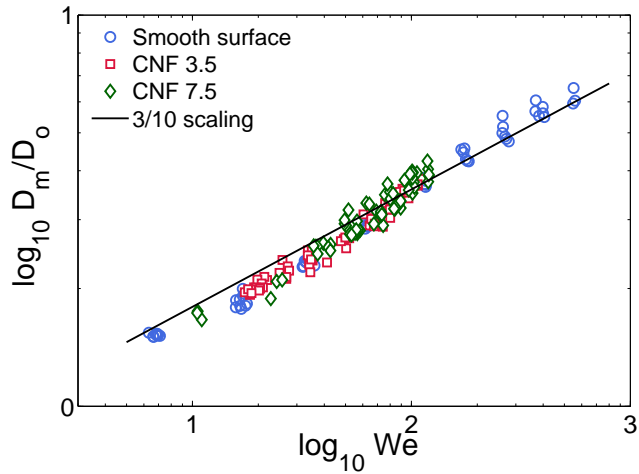} 
\caption{\small{Spreading factor $D_m/D$ for impacting FC-72 droplets on 
three surfaces: smooth silicon, and silicon covered by a 3.5 {\textmu}m-
and a  7.5 {\textmu}m-thick layer of carbon nanofibers. 
All the data points were obtained for impacts  in the film boiling regime 
for which the impacting drop did not fragment in smaller droplets. 
The solid line represents the scaling relation $D_m/D\sim {\rm We}^{3/10}$
derived by taking the vapor flow as the major driving mechanism for the 
spreading of the liquid\cite{tran13}. 
}} 
\label{spreading} 
\end{center} 
\end{figure}

We devote this section to quantifying the spreading factor of impacting droplets
in the film boiling regime. The spreading factor is defined as $D_m/D$, where 
$D_m$ is the maximum spreading diameter. 
In fig.\ \ref{spreading}, we show a log-log plot of $D_m/D$ versus ${\rm We}$ for 
all the impact experiments obtained on smooth and CNF surfaces. All the data 
points were collected for impacts in the film boiling regime and in the course 
of which the droplets did not disintegrate during the expanding 
phase. The Weber number ranges from 5 to 600. All data sets collected from the 
three different surfaces collapse on the same curve, showing that the 
spreading dynamics does not depend on the features and temperature of the 
surfaces. This result is consistent with the recent study of impacting 
droplets on micro-structured surfaces\cite{tran13}, which showed that the 
spreading factor is independent of the microstructures and depends very 
weakly on the surface temperature. Moreover, the spreading factor is in 
agreement with the scaling $D_m/D \propto {\rm We}^{3/10}$. 
This scaling law embodies the main assumption that the spreading 
of the liquid is driven by the vapor flow underneath the droplet 
\cite{tran13}. 
As a result, we conclude that the presence of the carbon nanofibers only
changes the transition temperature to film boiling of the impacting droplets,
but does not affect the dynamics of the vapor flow in the film boiling regime 
or the liquid spreading. 

 \section{Conclusions}
 We have explored the phase space $({\rm We}, T)$ of impact of FC-72 droplets 
on  heated smooth silicon surfaces and surfaces coated with nanofibers (CNF)  
of different length. Unexpectedly, we have found that the dynamic Leidenfrost 
temperature $T_L$, i.e., the transition temperature between  the contact and 
film boiling regimes, is higher on the CNF surfaces than on smooth silicon 
surface. Increasing the fiber length from $3.5\,$\textmu m to 
 $7.5\,$\textmu m causes $T_L$ to increase significantly due to the 
small time scale with which the nanofibers cool to the temperature of the 
vapor generated by the approaching liquid. Thus, the temperature of the 
fibers when contact with the liquid is established is much lower than their 
initial temperature. In other words,  the temperature of the 
CNF surfaces has to be set higher  than in the case of smooth silicon surfaces
 to bring the impact into film boiling regime. 
In contrast, the silicon microstructured surfaces studied in Ref.\cite{tran13}
maintain their temperature during impact and $T_L$ is lower, possibly because 
the liquid surface area which generates the vapor is larger due to the 
curvature caused by the micro-pillars.

In spite of the effect on $T_L$, we have found that, as long as the impact 
is in the film boiling regime, the spreading factor 
   of the droplet does not depend on whether the surface is smooth or 
   covered with carbon nanofibers, nor does it depend on the surface 
temperature. The spreading factor is consistent with the scaling law 
   $D_m/D \propto {\rm We}^{3/10}$, which was derived based on 
   the effect of vapor flow on the spreading dynamics\cite{tran13}.
   
   The increase in the dynamic Leidenfrost temperature 
   caused by nanofibers fabricated on silicon surfaces 
   has a considerable implication for various applications
   that require high operating temperature because CNF surfaces can 
operate at higher $T_L$ while still maintaining contact with the liquid. 

      
 
\acknowledgement
This study was financially supported by the European Research Council ERC \& FOM.
We gratefully acknowledge Dr. Roald M. Tiggelaar and 
Stefan Schlautmann of MCS group for their assistance in 
the fabrication of Ni-Ta substrates for the CNF synthesis,
M. Smithers (MESA+ Nanolab) for SEM imaging, 
B. Geerdink and Ruben Lubkemman for technical 
support.


\begin{mcitethebibliography}{34}
\providecommand*\natexlab[1]{#1}
\providecommand*\mciteSetBstSublistMode[1]{}
\providecommand*\mciteSetBstMaxWidthForm[2]{}
\providecommand*\mciteBstWouldAddEndPuncttrue
  {\def\EndOfBibitem{\unskip.}}
\providecommand*\mciteBstWouldAddEndPunctfalse
  {\let\EndOfBibitem\relax}
\providecommand*\mciteSetBstMidEndSepPunct[3]{}
\providecommand*\mciteSetBstSublistLabelBeginEnd[3]{}
\providecommand*\EndOfBibitem{}
\mciteSetBstSublistMode{f}
\mciteSetBstMaxWidthForm{subitem}{(\alph{mcitesubitemcount})}
\mciteSetBstSublistLabelBeginEnd
  {\mcitemaxwidthsubitemform\space}
  {\relax}
  {\relax}

\bibitem[Kim(2007)]{kim07}
Kim,~J. \emph{Int. J. Heat Fluid Flow} \textbf{2007}, \emph{28}, 753 --
  767\relax
\mciteBstWouldAddEndPuncttrue
\mciteSetBstMidEndSepPunct{\mcitedefaultmidpunct}
{\mcitedefaultendpunct}{\mcitedefaultseppunct}\relax
\EndOfBibitem
\bibitem[Agostini et~al.(2007)Agostini, Fabbri, Park, Wojtan, Thome, and
  Michel]{agostini07}
Agostini,~B.; Fabbri,~M.; Park,~J.~E.; Wojtan,~L.; Thome,~J.~R.; Michel,~B.
  \emph{Heat Transfer Eng.} \textbf{2007}, \emph{28}, 258--281\relax
\mciteBstWouldAddEndPuncttrue
\mciteSetBstMidEndSepPunct{\mcitedefaultmidpunct}
{\mcitedefaultendpunct}{\mcitedefaultseppunct}\relax
\EndOfBibitem
\bibitem[Ebadian and Lin(2011)Ebadian, and Lin]{EbadianLin11}
Ebadian,~M.~A.; Lin,~C.~X. \emph{J. Heat Transfer} \textbf{2011}, \emph{133},
  110801\relax
\mciteBstWouldAddEndPuncttrue
\mciteSetBstMidEndSepPunct{\mcitedefaultmidpunct}
{\mcitedefaultendpunct}{\mcitedefaultseppunct}\relax
\EndOfBibitem
\bibitem[Pautsch and Shedd(2005)Pautsch, and Shedd]{pautsch05}
Pautsch,~A.~G.; Shedd,~T.~A. \emph{Int. J. Heat Mass Transfer} \textbf{2005},
  \emph{48}, 3167--3175\relax
\mciteBstWouldAddEndPuncttrue
\mciteSetBstMidEndSepPunct{\mcitedefaultmidpunct}
{\mcitedefaultendpunct}{\mcitedefaultseppunct}\relax
\EndOfBibitem
\bibitem[Visaria and Mudawar(2009)Visaria, and Mudawar]{VisariaMudawar09}
Visaria,~M.; Mudawar,~I. \emph{IEEE Trans. Compon. Packag. Technol.}
  \textbf{2009}, \emph{32}, 784--793\relax
\mciteBstWouldAddEndPuncttrue
\mciteSetBstMidEndSepPunct{\mcitedefaultmidpunct}
{\mcitedefaultendpunct}{\mcitedefaultseppunct}\relax
\EndOfBibitem
\bibitem[Arcoumanis et~al.(1998)Arcoumanis, Cutter, and
  Whitelaw]{Arcoumanisetal98}
Arcoumanis,~C.; Cutter,~P.; Whitelaw,~D.~S. \emph{Chem. Eng. Res. Des.}
  \textbf{1998}, \emph{76}, 124--132\relax
\mciteBstWouldAddEndPuncttrue
\mciteSetBstMidEndSepPunct{\mcitedefaultmidpunct}
{\mcitedefaultendpunct}{\mcitedefaultseppunct}\relax
\EndOfBibitem
\bibitem[Panao and Moreira(2009)Panao, and Moreira]{PanaoMoreira09}
Panao,~M. R.~O.; Moreira,~A. L.~N. \emph{Int. J. Thermal Sci.} \textbf{2009},
  \emph{48}, 1853--1862\relax
\mciteBstWouldAddEndPuncttrue
\mciteSetBstMidEndSepPunct{\mcitedefaultmidpunct}
{\mcitedefaultendpunct}{\mcitedefaultseppunct}\relax
\EndOfBibitem
\bibitem[Mascarenhas and Mudawar(2012)Mascarenhas, and
  Mudawar]{MascarenhasMudawar12}
Mascarenhas,~N.; Mudawar,~I. \emph{Int. J. Heat Mass Transfer} \textbf{2012},
  \emph{55}, 2953--2964\relax
\mciteBstWouldAddEndPuncttrue
\mciteSetBstMidEndSepPunct{\mcitedefaultmidpunct}
{\mcitedefaultendpunct}{\mcitedefaultseppunct}\relax
\EndOfBibitem
\bibitem[Yoon et~al.(2010)Yoon, Figueroa, Brown, and Blanchat]{Yoonetal10}
Yoon,~S.~S.; Figueroa,~V.; Brown,~A.~L.; Blanchat,~T.~K. \emph{J. Fire Sci.}
  \textbf{2010}, \emph{28}, 109--139\relax
\mciteBstWouldAddEndPuncttrue
\mciteSetBstMidEndSepPunct{\mcitedefaultmidpunct}
{\mcitedefaultendpunct}{\mcitedefaultseppunct}\relax
\EndOfBibitem
\bibitem[Chen and Wang(2011)Chen, and Wang]{ChenWang11}
Chen,~P.-P.; Wang,~X.-S. \emph{Int. J. Heat Mass Transfer} \textbf{2011},
  \emph{54}, 4143--4147\relax
\mciteBstWouldAddEndPuncttrue
\mciteSetBstMidEndSepPunct{\mcitedefaultmidpunct}
{\mcitedefaultendpunct}{\mcitedefaultseppunct}\relax
\EndOfBibitem
\bibitem[Moreira et~al.(2010)Moreira, Moita, and Pan{\~a}o]{moreira10}
Moreira,~A. L.~N.; Moita,~A.~S.; Pan{\~a}o,~M.~R. \emph{Prog. Energy Combust.
  Sci.} \textbf{2010}, \emph{36}, 554--580\relax
\mciteBstWouldAddEndPuncttrue
\mciteSetBstMidEndSepPunct{\mcitedefaultmidpunct}
{\mcitedefaultendpunct}{\mcitedefaultseppunct}\relax
\EndOfBibitem
\bibitem[Berberovic et~al.(2011)Berberovic, Roisman, Jakirlic, and
  Tropea]{Berberovicetal11}
Berberovic,~E.; Roisman,~I.~V.; Jakirlic,~S.; Tropea,~C. \emph{Int. J. Heat
  Fluid Flow} \textbf{2011}, \emph{32}, 785--795\relax
\mciteBstWouldAddEndPuncttrue
\mciteSetBstMidEndSepPunct{\mcitedefaultmidpunct}
{\mcitedefaultendpunct}{\mcitedefaultseppunct}\relax
\EndOfBibitem
\bibitem[Yarin(2006)]{Yarin06}
Yarin,~A. \emph{Annu. Rev. Fluid Mech.} \textbf{2006}, \emph{38},
  159--192\relax
\mciteBstWouldAddEndPuncttrue
\mciteSetBstMidEndSepPunct{\mcitedefaultmidpunct}
{\mcitedefaultendpunct}{\mcitedefaultseppunct}\relax
\EndOfBibitem
\bibitem[Herbert et~al.(2013)Herbert, Gambaryan-Roisman, and
  Stephan]{Herbertetal13}
Herbert,~S.; Gambaryan-Roisman,~T.; Stephan,~P. \emph{Colloids Surf. A}
  \textbf{2013}, \emph{432}, 57--63\relax
\mciteBstWouldAddEndPuncttrue
\mciteSetBstMidEndSepPunct{\mcitedefaultmidpunct}
{\mcitedefaultendpunct}{\mcitedefaultseppunct}\relax
\EndOfBibitem
\bibitem[Bernardin et~al.(1997)Bernardin, Stebbins, and Mudawar]{bernardin97}
Bernardin,~J.~D.; Stebbins,~C.~J.; Mudawar,~I. \emph{Int. J. Heat Mass
  Transfer} \textbf{1997}, \emph{40}, 247--267\relax
\mciteBstWouldAddEndPuncttrue
\mciteSetBstMidEndSepPunct{\mcitedefaultmidpunct}
{\mcitedefaultendpunct}{\mcitedefaultseppunct}\relax
\EndOfBibitem
\bibitem[Tran et~al.(2013)Tran, Staat, Susarrey-Arce, Foertsch, van Houselt,
  Gardeniers, Prosperetti, Lohse, and Sun]{tran13}
Tran,~T.; Staat,~H. J.~J.; Susarrey-Arce,~A.; Foertsch,~T.~C.; van Houselt,~A.;
  Gardeniers,~J. G.~E.; Prosperetti,~A.; Lohse,~D.; Sun,~C. \emph{Soft Matter}
  \textbf{2013}, \emph{9}, 3272--3282\relax
\mciteBstWouldAddEndPuncttrue
\mciteSetBstMidEndSepPunct{\mcitedefaultmidpunct}
{\mcitedefaultendpunct}{\mcitedefaultseppunct}\relax
\EndOfBibitem
\bibitem[Bernardin and Mudawar(1999)Bernardin, and Mudawar]{bernardin99}
Bernardin,~J.~D.; Mudawar,~I. \emph{J. Heat Transfer} \textbf{1999},
  \emph{121}, 894--903\relax
\mciteBstWouldAddEndPuncttrue
\mciteSetBstMidEndSepPunct{\mcitedefaultmidpunct}
{\mcitedefaultendpunct}{\mcitedefaultseppunct}\relax
\EndOfBibitem
\bibitem[Wang et~al.(2000)Wang, Lin, and Chen]{wang00}
Wang,~A.~B.; Lin,~C.~H.; Chen,~C.~C. \emph{Phys. Fluids} \textbf{2000},
  \emph{12}, 1622\relax
\mciteBstWouldAddEndPuncttrue
\mciteSetBstMidEndSepPunct{\mcitedefaultmidpunct}
{\mcitedefaultendpunct}{\mcitedefaultseppunct}\relax
\EndOfBibitem
\bibitem[Bernardin and Mudawar(2004)Bernardin, and Mudawar]{bernardin04}
Bernardin,~J.~D.; Mudawar,~I. \emph{J. Heat Transfer} \textbf{2004},
  \emph{126}, 272--278\relax
\mciteBstWouldAddEndPuncttrue
\mciteSetBstMidEndSepPunct{\mcitedefaultmidpunct}
{\mcitedefaultendpunct}{\mcitedefaultseppunct}\relax
\EndOfBibitem
\bibitem[Tran et~al.(2012)Tran, Staat, Prosperetti, Sun, and Lohse]{tran12}
Tran,~T.; Staat,~H. J.~J.; Prosperetti,~A.; Sun,~C.; Lohse,~D. \emph{Phys. Rev.
  Lett.} \textbf{2012}, \emph{108}, 036101\relax
\mciteBstWouldAddEndPuncttrue
\mciteSetBstMidEndSepPunct{\mcitedefaultmidpunct}
{\mcitedefaultendpunct}{\mcitedefaultseppunct}\relax
\EndOfBibitem
\bibitem[Lee et~al.(2001)Lee, Kim, and Kiger]{Leeetal01}
Lee,~J.; Kim,~J.; Kiger,~K. \emph{Int. J. Heat Fluid Flow} \textbf{2001},
  \emph{22}, 188--223\relax
\mciteBstWouldAddEndPuncttrue
\mciteSetBstMidEndSepPunct{\mcitedefaultmidpunct}
{\mcitedefaultendpunct}{\mcitedefaultseppunct}\relax
\EndOfBibitem
\bibitem[Weickgenannt et~al.(2011)Weickgenannt, Zhang, Sinha-Ray, Roisman,
  Gambaryan-Roisman, Tropea, and Yarin]{weickgenannt11}
Weickgenannt,~C.~M.; Zhang,~Y.; Sinha-Ray,~S.; Roisman,~I.~V.;
  Gambaryan-Roisman,~T.; Tropea,~C.; Yarin,~A.~L. \emph{Phys. Rev. E}
  \textbf{2011}, \emph{84}, 036310\relax
\mciteBstWouldAddEndPuncttrue
\mciteSetBstMidEndSepPunct{\mcitedefaultmidpunct}
{\mcitedefaultendpunct}{\mcitedefaultseppunct}\relax
\EndOfBibitem
\bibitem[Weickgenannt et~al.(2011)Weickgenannt, Zhang, Lembach, Roisman,
  Gambaryan-Roisman, Yarin, and Tropea]{weickgenannt11b}
Weickgenannt,~C.~M.; Zhang,~Y.; Lembach,~A.~N.; Roisman,~I.~V.;
  Gambaryan-Roisman,~T.; Yarin,~A.~L.; Tropea,~C. \emph{Phys. Rev. E}
  \textbf{2011}, \emph{83}, 036305\relax
\mciteBstWouldAddEndPuncttrue
\mciteSetBstMidEndSepPunct{\mcitedefaultmidpunct}
{\mcitedefaultendpunct}{\mcitedefaultseppunct}\relax
\EndOfBibitem
\bibitem[Chen et~al.(2007)Chen, Chiu, and Lin]{chen07}
Chen,~R.~H.; Chiu,~S.~L.; Lin,~T.~H. \emph{Appl. Therm. Eng.} \textbf{2007},
  \emph{27}, 2079--2085\relax
\mciteBstWouldAddEndPuncttrue
\mciteSetBstMidEndSepPunct{\mcitedefaultmidpunct}
{\mcitedefaultendpunct}{\mcitedefaultseppunct}\relax
\EndOfBibitem
\bibitem[Chen et~al.(2007)Chen, Chiu, and Lin]{chen07b}
Chen,~R.-H.; Chiu,~S.-L.; Lin,~T.-H. \emph{Exp. Therm. Fluid Sci.}
  \textbf{2007}, \emph{32}, 587--595\relax
\mciteBstWouldAddEndPuncttrue
\mciteSetBstMidEndSepPunct{\mcitedefaultmidpunct}
{\mcitedefaultendpunct}{\mcitedefaultseppunct}\relax
\EndOfBibitem
\bibitem[Jun et~al.(2013)Jun, Sinha-Ray, and Yarin]{jun2013pool}
Jun,~S.; Sinha-Ray,~S.; Yarin,~A.~L. \emph{Int. J. Heat Mass Transfer}
  \textbf{2013}, \emph{62}, 99--111\relax
\mciteBstWouldAddEndPuncttrue
\mciteSetBstMidEndSepPunct{\mcitedefaultmidpunct}
{\mcitedefaultendpunct}{\mcitedefaultseppunct}\relax
\EndOfBibitem
\bibitem[Bitter(2010)]{bitter10}
Bitter,~J.~H. \emph{J. Mater. Chem.} \textbf{2010}, \emph{20}, 7312--7321\relax
\mciteBstWouldAddEndPuncttrue
\mciteSetBstMidEndSepPunct{\mcitedefaultmidpunct}
{\mcitedefaultendpunct}{\mcitedefaultseppunct}\relax
\EndOfBibitem
\bibitem[Nair et~al.(2013)Nair, Tiggelaar, Thakur, Gardeniers, van Houselt, and
  Lefferts]{nair12}
Nair,~H.; Tiggelaar,~R.~M.; Thakur,~D.~B.; Gardeniers,~J. G.~E.; van
  Houselt,~A.; Lefferts,~L. \emph{Chem. Eng. J.} \textbf{2013}, \emph{227},
  56--65\relax
\mciteBstWouldAddEndPuncttrue
\mciteSetBstMidEndSepPunct{\mcitedefaultmidpunct}
{\mcitedefaultendpunct}{\mcitedefaultseppunct}\relax
\EndOfBibitem
\bibitem[Fujimoto et~al.(2010)Fujimoto, Oku, Ogihara, and Takuda]{fujimoto10}
Fujimoto,~H.; Oku,~Y.; Ogihara,~T.; Takuda,~H. \emph{Int. J. Multiphase Flow}
  \textbf{2010}, \emph{36}, 620--642\relax
\mciteBstWouldAddEndPuncttrue
\mciteSetBstMidEndSepPunct{\mcitedefaultmidpunct}
{\mcitedefaultendpunct}{\mcitedefaultseppunct}\relax
\EndOfBibitem
\bibitem[Qu{\'e}r{\'e}(2013)]{quere13}
Qu{\'e}r{\'e},~D. \emph{Annu. Rev. Fluid Mech.} \textbf{2013}, \emph{45},
  197--215\relax
\mciteBstWouldAddEndPuncttrue
\mciteSetBstMidEndSepPunct{\mcitedefaultmidpunct}
{\mcitedefaultendpunct}{\mcitedefaultseppunct}\relax
\EndOfBibitem
\bibitem[Incropera et~al.(2010)Incropera, DeWitt, Bergman, and
  Lavine]{Incroperaetal}
Incropera,~F.~P.; DeWitt,~D.~P.; Bergman,~T.~L.; Lavine,~A.~S.
  \emph{Fundamentals of Heat and Mass Transfer}, 7th ed.; Wiley, 2010\relax
\mciteBstWouldAddEndPuncttrue
\mciteSetBstMidEndSepPunct{\mcitedefaultmidpunct}
{\mcitedefaultendpunct}{\mcitedefaultseppunct}\relax
\EndOfBibitem
\bibitem[Mandre et~al.(2009)Mandre, Mani, and Brenner]{mandre09}
Mandre,~S.; Mani,~M.; Brenner,~M.~P. \emph{Phys. Rev. Lett.} \textbf{2009},
  \emph{102}, 134502\relax
\mciteBstWouldAddEndPuncttrue
\mciteSetBstMidEndSepPunct{\mcitedefaultmidpunct}
{\mcitedefaultendpunct}{\mcitedefaultseppunct}\relax
\EndOfBibitem
\bibitem[Mayhew and Prakash(2013)Mayhew, and Prakash]{MayhewPrakash13}
Mayhew,~E.; Prakash,~V. \emph{Carbon} \textbf{2013}, \emph{62}, 493--500\relax
\mciteBstWouldAddEndPuncttrue
\mciteSetBstMidEndSepPunct{\mcitedefaultmidpunct}
{\mcitedefaultendpunct}{\mcitedefaultseppunct}\relax
\EndOfBibitem
\end{mcitethebibliography}
\providecommand*\mcitethebibliography{\thebibliography}
\csname @ifundefined\endcsname{endmcitethebibliography}
  {\let\endmcitethebibliography\endthebibliography}{}

\end{document}